\documentclass[twocolumn,showpacs,aps,pre]{revtex4-1}

\usepackage{amssymb,amsmath}
\usepackage[dvips]{graphicx}

\begin{document}

\title{Monte-Carlo simulations of the clean and disordered contact process in three dimensions}

\author{Thomas Vojta}
\affiliation{Department of Physics, Missouri University of Science and Technology,
Rolla, MO 65409, USA}

\begin{abstract}
The absorbing-state transition in the three-dimensional contact process with and without quenched
randomness is investigated by means of Monte-Carlo simulations. In
the clean case, a reweighting technique is combined with a careful extrapolation of the data
to infinite time to determine with high accuracy the critical behavior in the three-dimensional
directed percolation universality class.
In the presence of quenched spatial disorder, our data demonstrate that the absorbing-state transition
is governed by an unconventional infinite-randomness critical point featuring activated
dynamical scaling. The critical behavior of this transition does not depend on the disorder strength,
i.e., it is universal. Close to the disordered critical point, the dynamics is characterized by
the nonuniversal power laws typical of a Griffiths phase. We compare our findings
to the results of other numerical methods, and we relate them to a general classification of phase
transitions in disordered systems based on the rare region dimensionality.
\end{abstract}

\date{\today}

\maketitle


\section{Introduction}
\label{sec:intro}

The macroscopic behavior of many-particle systems far from equilibrium can abruptly
change when an external parameter is changed. The resulting nonequilibrium phase
transitions separate different nonequilibrium steady states. They are characterized
by strong fluctuations and cooperative phenomena over large distances and times,
analogous to the behavior at equilibrium phase transitions. Examples of nonequilibrium
phase transitions can be found in catalytic reactions, growing interfaces, turbulence,
and traffic jams as well as in the dynamics of epidemics and other biological populations
(see, e.g., Refs.\
\cite{ZhdanovKasemo94,SchmittmannZia95,MarroDickman99,Hinrichsen00,Odor04,Luebeck04,TauberHowardVollmayrLee05, HenkelHinrichsenLuebeck_book08}).

A well-studied class of nonequilibrium phase transitions are the
absorbing-state transitions between active, fluctuating steady states and inactive,
absorbing states in which fluctuations cease completely. The generic universality class
for absorbing-state transitions is the directed percolation (DP) class
\cite{GrassbergerdelaTorre79}. Janssen and Grassberger \cite{Janssen81,Grassberger82}
conjectured that all absorbing-state transitions with a scalar order parameter and
short-range interactions belong to this class, provided they do not feature
extra symmetries or conservation laws.
Additional symmetries or conservation laws can lead to  other universality classes such as
the parity conserving class or $Z_2$-symmetric directed percolation (DP2)
(see, e.g., Refs.\ \cite{Hinrichsen00,Odor04}).

Although absorbing state transitions are ubiquitous in theory and computer simulations,
experimental observations of their universality classes were lacking for a
long time \cite{Hinrichsen00b}. A full verification of the DP universality class
was recently achieved in the transition between two turbulent states in a liquid crystal
\cite{TKCS07}. Other absorbing state transitions were found in periodically driven
suspensions \cite{CCGP08,FFGP11} and in superconducting vortices \cite{OkumaTsugawaMotohashi11}.

In many experimental systems, one can expect impurities and defects to play an important
role. Indeed, it has been suggested \cite{Hinrichsen00b} that such quenched spatial disorder
is one of the key reasons for the surprising rarity of the DP universality class
in experiments. The influence of disorder on absorbing state
transitions is therefore a prime problem in the field. According to the
Harris criterion \cite{Harris74}, a clean critical point is stable against the
introduction of weak spatial disorder if its correlation length critical exponent $\nu_\perp$
fulfills the inequality $d\nu_\perp > 2$ where $d$ is the space dimensionality.
The values of $\nu_\perp$ in the clean DP universality class are approximately 1.1 in one dimension,
0.73 in two dimensions, and 0.58 in three dimensions \cite{Hinrichsen00}. The Harris criterion is
thus violated, and spatial disorder is expected to change the critical behavior.
This heuristic result was confirmed by a field-theoretic renormalization group
study \cite{Janssen97} which found runaway flow towards large disorder. Early Monte-Carlo
simulations
\cite{Noest86,Noest88,BramsonDurrettSchonmann91,MoreiraDickman96,DickmanMoreira98,WACH98,CafieroGabrielliMunoz98}
demonstrated unusually slow dynamics but could not resolve the ultimate fate of the
transition.

In recent years, a comprehensive understanding of the one-dimensional disordered contact
process has been achieved by a combination of analytical and numerical approaches.
A strong-disorder renormalization group (SDRG) analysis \cite{HooyberghsIgloiVanderzande03,HooyberghsIgloiVanderzande04}
established that the critical point is of exotic infinite-randomness type and
characterized by activated (exponential) dynamical scaling. It belongs to
the same universality class as the random transverse-field Ising chain \cite{Fisher92,Fisher95},
at least for sufficiently strong disorder. These predictions were confirmed by
large-scale Monte-Carlo simulations \cite{VojtaDickison05} that also provided evidence
for the critical behavior being universal, i.e., independent of the disorder strength.
In higher dimensions, the SDRG cannot be solved analytically. However, by using
a numerical implementation of the SDRG \cite{MMHF00}, the infinite-randomness scenario
was found to be valid in two dimensions, in agreement with Monte-Carlo simulations of
the contact process on diluted lattices \cite{OliveiraFerreira08,VojtaFarquharMast09}
\footnote{Somewhat surprisingly, the contact process on a two-dimensional random
Voronoi triangulation \protect{\cite{OAFD08}} appears to show the clean DP critical behavior, in contradiction to
the Harris criterion. A similar result was also obtained for an Ising model
on a Voronoi triangulation \protect{\cite{JankeVillanova02}}. The reasons for these contradictions
are not understood so far, possibly the Voronoi triangulation implements rather
weak disorder.}.

Here, we extend our Monte-Carlo simulations of the contact process on diluted lattices to three
space dimensions. Using large lattices of up to $999^3$ sites and long times up to $10^8$, we
provide strong evidence for the non-equilibrium phase transition of the
disordered contact process being governed by an infinite-randomness critical point. We determine
the critical exponents and find them to be universal, i.e., independent of disorder strength.
In contrast, the dynamics in the Griffiths region between the clean and disordered critical points
is characterized by nonuniversal power laws. As a byproduct
of our simulations, we also obtain high-precision estimates for the critical exponents of the
clean contact process in three dimensions.

The paper is organized as follows. The contact process on a diluted lattice is introduced in Sec.\
\ref{sec:cp}. We briefly summarize the scaling theories for conventional and infinite-randomness
critical points in Sec.\ \ref{sec:scaling}. In Sec.\ \ref{sec:MC}, we describe our simulation method
and present the results. We conclude in Sec.\ \ref{sec:conclusions}.

\section{Definition of the contact process}
\label{sec:cp}

The contact process \cite{HarrisTE74} can be viewed as a model for the spreading of an
epidemic in space. Consider a hypercubic $d$-dimensional lattice of $L^d$  sites.
Each site can be in one of two states,
either active (infected) or inactive (healthy).
The time evolution of the contact process is a continuous-time Markov process
during which infected sites heal spontaneously at a rate $\mu$ while
healthy sites become infected by their neighbors at a rate $\lambda n /(2d)$.
Here, $n$ is the number of sick nearest neighbors of the given site.
The infection rate $\lambda$ and the healing rate $\mu$ (which can be set to unity)
are the external control parameters that govern the behavior of the system.

The steady  states of the contact process can be easily understood at a qualitative
level. For $\lambda \ll \mu$, healing dominates over infection, and the epidemic eventually dies
out completely. The system therefore always ends up in the absorbing steady state without
any infected sites. This is the inactive phase. In contrast,
the density of infected sites remains nonzero in the long-time limit if the
infection rate $\lambda$ is sufficiently large, i.e., the system is in the active phase.
The nonequilibrium transition between these two phases, which occurs
at a critical infection rate $\lambda_c^0$, belongs to the DP universality
class.

\begin{figure}[tb]
\centerline{\includegraphics[width=8.5cm]{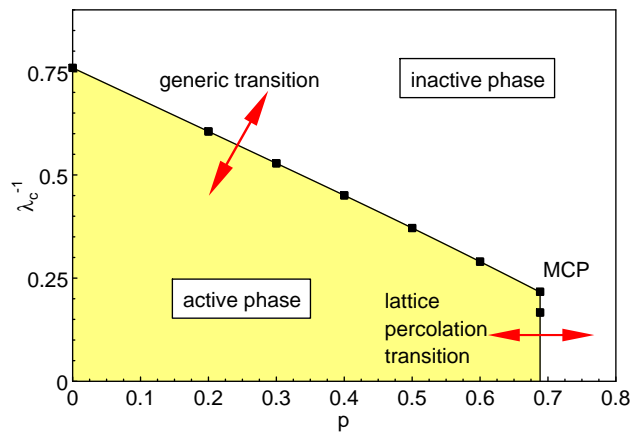}}
\caption{(Color online) Phase diagram of the contact
process on a site-diluted cubic lattice
         (inverse critical infection rate $\lambda_c^{-1}$ vs vacancy concentration $p$).
         MCP marks the multicritical point. The
          black dots show the actual simulation results, the lines are guides to
          the eye.}
\label{fig:pd}
\end{figure}

Quenched spatial disorder can be introduced into the contact process in different ways,
e.g., by making the infection and healing rates random variables, or by using a random
lattice instead of a regular one. Here, we randomly dilute the regular lattice
by removing each site with probability $p$ \footnote{We define $p$ is the
fraction of sites removed rather than the fraction of sites present.}. In the context of an
epidemic, a vacancy can be interpreted as a site that is immune
against the infection. For vacancy concentrations $p$ below the
percolation threshold $p_c$, the lattice still has an infinite connected cluster
of sites that can support an active phase of the contact process.
If the vacancy concentration is above $p_c$, an infinite cluster does not exist.
Instead, the lattice consists of disconnected finite-size clusters. As the epidemic
dies out on any finite cluster in the long-time limit, an active phase is impossible
for $p>p_c$. This leads to the phase diagram shown in Fig.\ \ref{fig:pd}.
Specifically, there are two different nonequilibrium phase transitions: (i) the so-called
generic transition for $p<p_c$, driven by the dynamic fluctuations of the contact
process and (ii) the lattice percolation transition occurring at $p=p_c$ for sufficiently
large infection rates \cite{VojtaLee06,LeeVojta09}.
The two phase transition lines meet at a multicritical point
\footnote{ In two space dimensions, the multicritical point was
studied in Ref.\ \protect{\cite{DahmenSittlerHinrichsen07}}.}.

The central quantity of the contact process is the density of infected sites
at time $t$,
\begin{equation}
\rho(t) = \frac 1 {L^d} \sum_{\mathbf{r}} \langle  n_\mathbf{r}(t) \rangle~.
\label{eq:rho_definition}
\end{equation}
Here, $n_\mathbf{r}(t)$ is the occupation of site $\mathbf{r}$ at time $t$, i.e.,
$n_\mathbf{r}(t)=1$ if the site is infected and $n_\mathbf{r}(t)=0$
if it is healthy.  $\langle \ldots \rangle$ denotes the
average over all
realizations of the Markov process.
The order parameter of the absorbing-state phase transition is given by the steady
state density
\begin{equation}
\rho_{\rm stat} = \lim_{t\to\infty} \rho(t)~.
\label{eq:OP_definition}
\end{equation}

\section{Scaling theories of absorbing state transitions}
\label{sec:scaling}

In this section, we summarize the scaling theories of the nonequilibrium transitions
in the clean and disordered contact process to
 the extent necessary for analyzing our Monte-Carlo data. We contrast the cases of
conventional power-law scaling and activated scaling.
More details can be found, for instance, in Ref.\ \cite{Hinrichsen00} for the
power-law case and in Refs.\ \cite{Vojta06,IgloiMonthus05} for the activated case.

\subsection{Conventional critical points}
\label{subsec:power-law}

The DP universality class is characterized by three independent critical exponents, for example,
$\beta$, $\nu_\perp$, and $z$. The order parameter exponent $\beta$ describes how the steady state density varies
as the infection rate $\lambda$ approaches its critical value $\lambda_c$ from above,
\begin{equation}
\rho_{\rm stat} \sim (\lambda-\lambda_c)^\beta \sim \Delta^\beta~.
\label{eq:beta}
\end{equation}
Here, $\Delta=(\lambda-\lambda_c)/\lambda_c$ is the dimensionless distance from criticality.
The correlation length exponent $\nu_\perp$ describes the
divergence of the correlation length $\xi_\perp$ at criticality,
\begin{equation}
\xi_\perp \sim |\Delta|^{-\nu_{\perp}}~.
\label{eq:nu}
\end{equation}
The correlation time $\xi_\parallel$ diverges like a power of the correlation length,
\begin{equation}
\xi_\parallel \sim \xi_\perp^z, \label{eq:powerlawscaling}
\end{equation}
which defines the dynamical exponent $z$. In terms of these exponents, the scaling form
of the density as a function of $\Delta$, time $t$, and system size $L$ reads
\begin{equation}
\rho(\Delta,t,L) = b^{\beta/\nu_\perp} \rho(\Delta b^{-1/\nu_\perp},t b^z, L b)~.
\label{eq:rho}
\end{equation}
Here, $b$ is an arbitrary dimensionless length scale factor.

If the time evolution starts at time 0 from a single infected site in an otherwise inactive lattice,
one can ask what is the probability that an active cluster survives at time $t$. In the DP universality
class, this survival probability $P_s$ has the same scaling form as the density
\footnote{At general absorbing state transitions, e.g., with several
    absorbing states, the survival probability scales with an exponent $\beta'$ which may
    be different from $\beta$ (see, e.g., \protect{\cite{Hinrichsen00}}).},
\begin{equation}
P_s(\Delta,t,L) = b^{\beta/\nu_\perp} P_s(\Delta b^{-1/\nu_\perp},t b^z, L b)~.
\label{eq:Ps}
\end{equation}
The correlation (or pair connectedness) function $C(\mathbf{r},t)=\langle
n_{\mathbf{r}}(t) \, n_{0}(0) \rangle$ is given by the probability that site
$\mathbf{r}$ is infected at time $t$ when the time evolution starts from a single infected site at
$\mathbf{r}=0$ and time $0$. The scale dimension of $C$ is $2\beta/\nu_\perp$
because it involves a product of two densities, leading to the  scaling form
 \footnote{This relation relies on hyperscaling; it is only valid below the
    upper critical dimension $d_c^+$, which is four for directed percolation}
\begin{equation}
C(\Delta,\mathbf{r},t,L) = b^{2\beta/\nu_\perp} C(\Delta b^{-1/\nu_\perp}, \mathbf{r}b, t
b^z, L b)~.
\label{eq:C}
\end{equation}
The total number $N_s$ of sites in the active cluster can be calculated
by integrating the correlation function over all space, resulting in
\begin{equation}
N_s(\Delta,t,L) = b^{2\beta/\nu_\perp - d} N_s(\Delta b^{-1/\nu_\perp},t b^z, L b)~.
\label{eq:Ns}
\end{equation}
The mean-square radius $R$ of the active cluster has the dimension of a length.
Its scaling form therefore reads
\begin{equation}
R(\Delta,t,L) = b^{-1} R(\Delta b^{-1/\nu_\perp},t b^z, L b)~.
\label{eq:R}
\end{equation}

The functional dependencies  of $\rho$, $P_s$, $N_s$ and $R$ on the parameters
$\Delta$, $t$, and $L$ can be easily derived from the scaling forms by setting the scale
factor $b$ to appropriate values. This leads to the following time dependencies at the critical point
$\Delta=0$ and in the thermodynamic limit $L\to \infty$.
In the long-time limit, the density of infected sites and the survival
probability obey the power laws
\begin{equation}
\rho(t) \sim t^{-\delta}, \qquad P_s(t) \sim t^{-\delta}
\label{eq:Ps_t}
\end{equation}
with $\delta=\beta/(\nu_\perp z)$. The mean-square radius and number of infected sites of a
cluster starting from a single seed site behave as
\begin{equation}
R(t) \sim t^{1/z}, \qquad N_s(t) \sim t^\Theta
\label{eq:Ns_t}
\end{equation}
where $\Theta=d/z - 2\beta/(\nu_\perp z)$ is the so-called critical initial slip
exponent. By taking the derivative of eqs.\ (\ref{eq:rho}), (\ref{eq:Ps}), and (\ref{eq:Ns}) with
respect to $\Delta$, we also find that
\begin{equation}
\frac{\partial \ln \rho}{\partial \Delta} \sim \frac{\partial \ln P_s}{\partial \Delta} \sim \frac{\partial \ln N_s}{\partial \Delta} \sim t^{1/(\nu_\perp z)}
\label{eq:dlnrho}
\end{equation}
which will be useful for measuring $\nu_\perp$.

\subsection{Infinite-randomness critical points}
\label{subsec:activated}

Infinite-randomness critical points feature extremely slow dynamics, represented
by an exponential (activated) relation between correlation length and time
\begin{equation}
\ln(\xi_\parallel/t_0) \sim \xi_\perp^\psi, \label{eq:activatedscaling}
\end{equation}
rather than the power-law dependence (\ref{eq:powerlawscaling}).
It is characterized by the so-called tunneling exponent $\psi$, and $t_0$ is a
nonuniversal microscopic time scale.
The exponential relation between time and length
scales implies that the dynamical exponent $z$ is formally infinite.
In contrast to the dynamical scaling, the static scaling relations
remain of power-law type, i.e., eqs.\ (\ref{eq:beta}) and (\ref{eq:nu}) remain valid.

The scaling forms of disorder-averaged observables can be obtained by
simply substituting the variable combination $\ln(t/t_0) b^\psi$  for $t b^z$
in the arguments of the scaling functions:
\begin{equation}
\rho(\Delta,\ln(t/t_0),L) = b^{\beta/\nu_\perp} \rho(\Delta b^{-1/\nu_\perp},\ln(t/t_0) b^\psi, L
b)~, \label{eq:rho_activated}
\end{equation}
\begin{equation}
P_s(\Delta,\ln(t/t_0),L) = b^{\beta/\nu_\perp} P_s(\Delta b^{-1/\nu_\perp},\ln(t/t_0) b^\psi,L
b)~, \label{eq:Ps_activated}
\end{equation}
\begin{equation}
 N(\Delta,\ln(t/t_0),L) = b^{2\beta/\nu_\perp -d} N(\Delta
b^{-1/\nu_\perp},\ln(t/t_0) b^\psi,L b)~,  \label{eq:Ns_activated}
\end{equation}
\begin{equation}
 R(\Delta,\ln(t/t_0),L) = b^{-1} R(\Delta
b^{-1/\nu_\perp},\ln(t/t_0) b^\psi,L b)~.  \label{eq:R_activated}
\end{equation}

Consequently, the critical time dependencies of the density of active sites and the survival probability
(in the thermodynamic limit) are logarithmic,
\begin{equation}
\rho(t) \sim [\ln(t/t_0)]^{-\bar\delta}, \qquad P_s(t) \sim [\ln(t/t_0)]^{-\bar\delta}
\label{eq:Ps_t_activated}
\end{equation}
with $\bar\delta=\beta/(\nu_\perp \psi)$. The radius and number of active sites in a
cluster starting from a single seed site vary as
\begin{equation}
R(t) \sim [\ln(t/t_0)]^{1/\psi}, \qquad N_s(t) \sim [\ln(t/t_0)]^{\bar\Theta}
\label{eq:Ns_t_activated}
\end{equation}
with $\bar\Theta=d/\psi-2\beta/(\nu_\perp \psi)$.
Taking the derivatives of eqs.\ (\ref{eq:rho_activated}), (\ref{eq:Ps_activated}), and (\ref{eq:Ns_activated}) with
respect to $\Delta$ yields
\begin{equation}
\frac{\partial \ln \rho}{\partial \Delta} \sim \frac{\partial \ln P_s}{\partial \Delta} \sim \frac{\partial \ln N_s}{\partial \Delta}
\sim [\ln(t/t_0)]^{1/(\nu_\perp \psi)}~.
\label{eq:dlnrho_activated}
\end{equation}

\subsection{Griffiths region}
\label{subsec:griffiths}

In the presence of spatial disorder,
the contact process displays unconventional behavior not just at the critical point but
also in its vicinity because rare active spatial regions dominate the long-time dynamics.
This phenomenon is an example of the well-known Griffiths singularities \cite{Griffiths69} that generally
occur at phase transitions in disordered systems (see Ref.\ \cite{Vojta06} for a review).
The Griffiths singularities in the
spatially disordered contact process can be understood as follows \cite{Noest86}.

The inactive phase must be divided into two regions. (i) If the infection rate is below the
clean critical value, $\lambda<\lambda_c^0$, the behavior is conventional. This means, the system
approaches the absorbing state exponentially fast. The decay time increases with
$\lambda$ and diverges as $|\lambda-\lambda_c^0|^{-z\nu_\perp}$ where $z$ and $\nu_\perp$
are the exponents of the clean critical point \cite{VojtaDickison05,DickisonVojta05}.

(ii) If the infection rate is in the so-called Griffiths region (or Griffiths phase)
between the clean and dirty critical values, $\lambda_c^0 < \lambda < \lambda_c$,
the system is globally still in the inactive phase (i.e., it eventually decays into the absorbing state).
However, in the thermodynamic limit, one can find arbitrarily large spatial regions devoid of vacancies.
These rare regions are locally in the active phase. Although they cannot support a
non-zero steady state density because they are of finite size, their time decay is very slow
as it requires a rare, exceptionally large density fluctuation.

The contribution of the rare regions to the density
of infected sites
can be expressed as the integral
\begin{equation}
\rho(t) \sim \int dL_r ~L_r^d ~w(L_r) \exp\left[-t/\tau(L_r)\right]~. \label{eq:rrevo}
\end{equation}
Here, $w$ denotes the probability for finding a spatial region of size $L_r$
that does not contain any vacancies, and $\tau(L_r)$ is the life time of the contact process
on such a rare region.  Basic combinatorics gives
\begin{equation}
 w(L_r) \sim \exp( -\tilde p L_r^d)
\end{equation}
with $\tilde p = - \ln(1-p)$ (up to pre-exponential factors).
In the Griffiths phase, the life time of a rare region depends exponentially on its volume,
\begin{equation}
\tau(L_r) \sim \exp(a L_r^d)
\end{equation}
because a coordinated fluctuation of the entire region is necessary to take it to the
absorbing state \cite{Noest86,Noest88,Schonmann85}. The constant $a$
vanishes at the clean critical infection rate $\lambda_c^0$ and increases with $\lambda$.
Evaluating the integral (\ref{eq:rrevo}) in saddle-point approximation, we
obtain a power-law time dependence for the density. The survival probability $P_s$ shows
exactly the same time dependence,
\begin{equation}
\rho(t) \sim P_s(t) \sim  t^{-\tilde p/a}  = t^{-d/z'}~,
\label{eq:griffithspower}
\end{equation}
where $z'=da/\tilde p$ is the \emph{nonuniversal} dynamical exponent in the
Griffiths region.
The behavior of $z'$ close to the dirty critical point can
be obtained from the SDRG analysis
\cite{HooyberghsIgloiVanderzande04,Fisher95,MMHF00}. As $\lambda$ approaches $\lambda_c$, $z'$ diverges as
\begin{equation}
z' \sim  |\lambda-\lambda_c|^{-\psi\nu_\perp}
\label{eq:z_prime}
\end{equation}
where $\psi$ and $\nu_\perp$ are the critical exponents of the infinite-randomness critical point.

\section{Monte-Carlo simulations}
\label{sec:MC}

\subsection{Simulation method}
\label{subsec:MC_method}

To perform Monte Carlo simulations of the contact process on randomly diluted cubic lattices,
we followed the implementation described, for instance, by Dickman \cite{Dickman99}. The algorithm starts at time $t=0$ from
some configuration of infected  and healthy sites and consists of a sequence of events. During each event
an infected site is randomly chosen from a list of all $N_a$ infected sites, then a process is selected,
either infection of a neighbor with probability $\lambda/(1+ \lambda)$ or healing with probability $1/(1+ \lambda)$.
For infection, one of the six neighbor sites is chosen at random. The infection succeeds if this neighbor
is healthy (and not a vacancy site). The time is then incremented by $1/N_a$.

Using this algorithm, we simulated systems with sizes of up to $999^3$ sites and vacancy
concentrations $p=0$, 0.2, 0.3, 0.4, 0.5, 0.6 and $p_c=0.6883920$ \cite{LorenzZiff98}. To cope with the
slow dynamics of the disordered contact process, we simulated long times up to $10^8$.
All results were averaged over a large number of disorder configurations, precise numbers will be given
below.

We carried out two different types of simulations. (i) The majority of runs
started from a single active site in an otherwise inactive
lattice (spreading runs); we monitored the survival probability $P_s(t)$, the number of sites $N_s(t)$
of the active cluster, and its radius $R(t)$. (ii) For comparison, we also performed
a few runs that started from a completely active lattice during which we observed the time evolution
of the density $\rho(t)$.

We employed two different high-quality, long-period random number generators. Most simulations
used LFSR113 proposed by L'Ecuyer \cite{Lecuyer99}. We verified the validity of the results by means of
the 2005 version of Marsaglia's KISS \cite{Marsaglia05}.
The total computational effort for the work described in this paper was about 100\,000 CPU days
on the Pegasus cluster at Missouri S\&T.

Figure \ref{fig:pd} gives an overview of the phase diagram
resulting from these simulations. As expected, the critical infection rate $\lambda_c$
increases with increasing impurity concentration.

\subsection{Contact process on an undiluted lattice}
\label{subsec:MC_clean}

The purpose of studying the clean three-dimensional contact process is two-fold, (i)
to test our implementation of the contact process and (ii) to compute highly accurate estimates of
the critical exponents in the three-dimensional DP universality class. To reduce the numerical
effort, we applied the clever reweighting technique proposed in Ref.\ \cite{Dickman99}.

After a few test calculations aimed at bracketing the critical point, we performed two large
spreading runs (starting from a single active  site) at $\lambda=1.3168400$. By reweighting with
a step $\Delta \lambda=0.0000025$, we generated data for $\lambda$ between 1.3168150 and
1.3168650. The first run consisted of $4\times 10^8$ trials using the LFSR113 random number
generator, the second consisted of $5\times 10^8$ trials using the KISS random number
generator. The maximum time of both runs was $5\times 10^4$. As the data of both runs agree within
their statistical error, we averaged their results. The system size, $850^3$ sites, was chosen
such that the active cluster stayed smaller than the sample during the entire time evolution,
eliminating finite-size effects.

\begin{figure}
\centerline{\includegraphics[width=8.5cm]{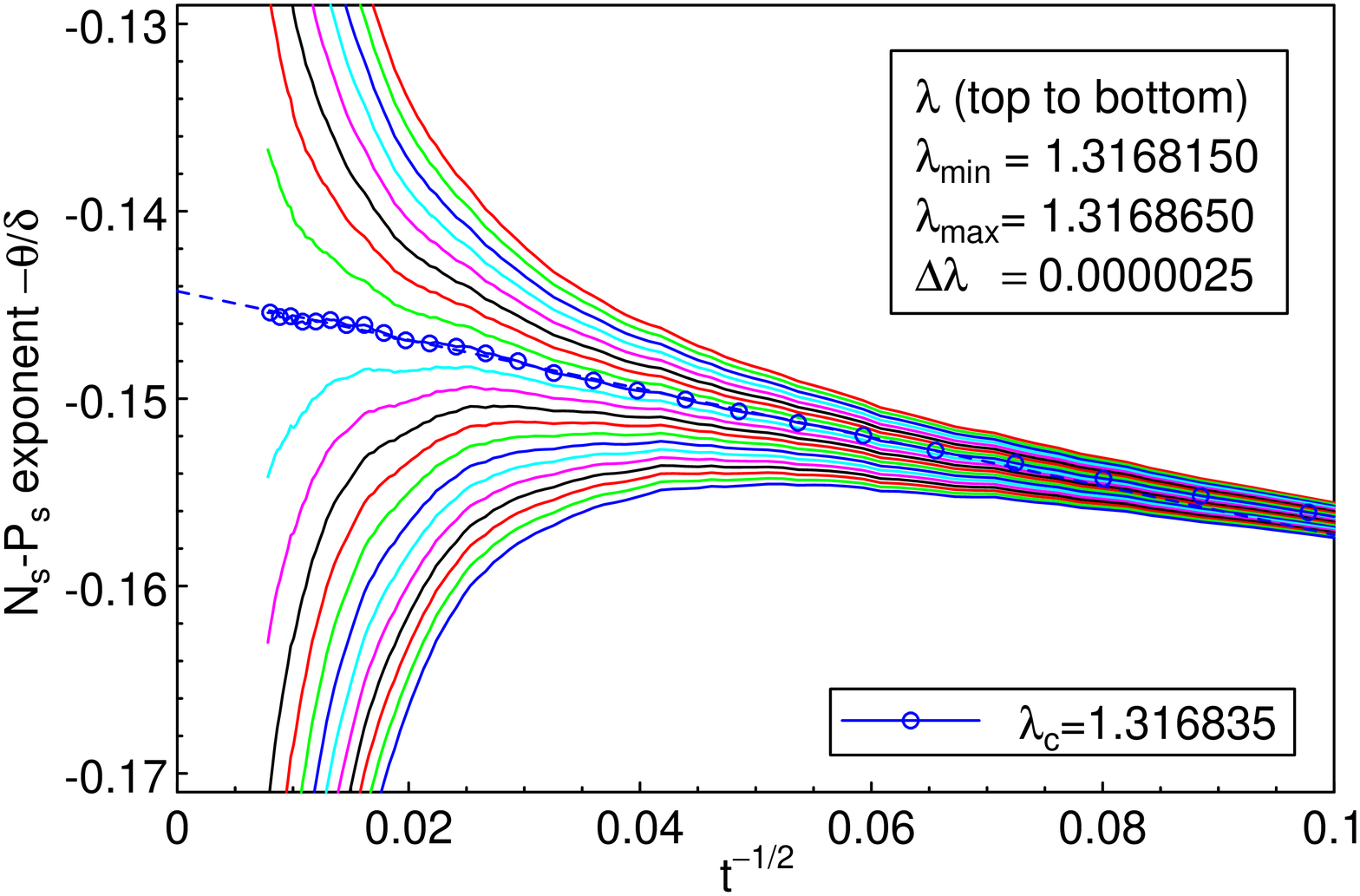}}
\caption{(Color online) Effective critical exponent $-\Theta/\delta = (d\ln N_s) / (d\ln P_s)$ vs.
         $t^{-1/2}$. The critical curve is marked by dots, and the dashed line is a linear
         extrapolation to $t=\infty$. }
\label{fig:cl_nsps_exponent}
\end{figure}

To find the location of the critical point and to measure the critical exponents, we define effective (running)
exponents via the logarithmic derivatives of various observables. These
effective exponents are then extrapolated to $t=\infty$. The finite-size scaling exponent $\beta/\nu_\perp$
(scale dimension of the order parameter) can be determined from the relation between $N_s$ and $P_s$.
Combining (\ref{eq:Ps_t}) and (\ref{eq:Ns_t}) yields $N_s \sim P_s^{-\Theta/\delta}$ with
$\Theta/\delta = 3\nu_\perp/\beta - 2$. Figure \ref{fig:cl_nsps_exponent} shows the effective
exponent $(d\ln N_s) /(d\ln P_s)$ as a function of $t^{-y}$ with $y=1/2$.
(The value 1/2 was chosen empirically to allow a linear extrapolation to $t=\infty$.) From this plot,
we estimate the critical infection rate to be
\begin{equation}
\lambda_c^0 = 1.316835 (1)~.
\label{eq:cl_lambda_c}
\end{equation}
We verified this value by performing an extra run directly at $\lambda=1.316835$ using $4\times 10^8$ trials on
a system of size $999^3$ with a maximum time of $10^5$.

Extrapolating the effective exponent to $t=\infty$, we obtain $\Theta/\delta=0.1442(3)_{\rm sys}(2)_{\rm ran}$
where the values in brackets represent estimates of the systematic and random errors of the last digit. The
systematic error stems from the uncertainties of $\lambda_c^0$ and the extrapolation exponent $y$ while
the random error is due to the Monte-Carlo noise. The resulting value of the finite-size scaling exponent is
$\beta/\nu_\perp = 1.3991(4)$. An estimate for this exponent can also be obtained from the relation between $N_s$ and $R$.
Extrapolating the effective exponent as above yields the identical value $\beta/\nu_\perp = 1.3991(4)$.

\begin{figure}
\centerline{\includegraphics[width=8.5cm]{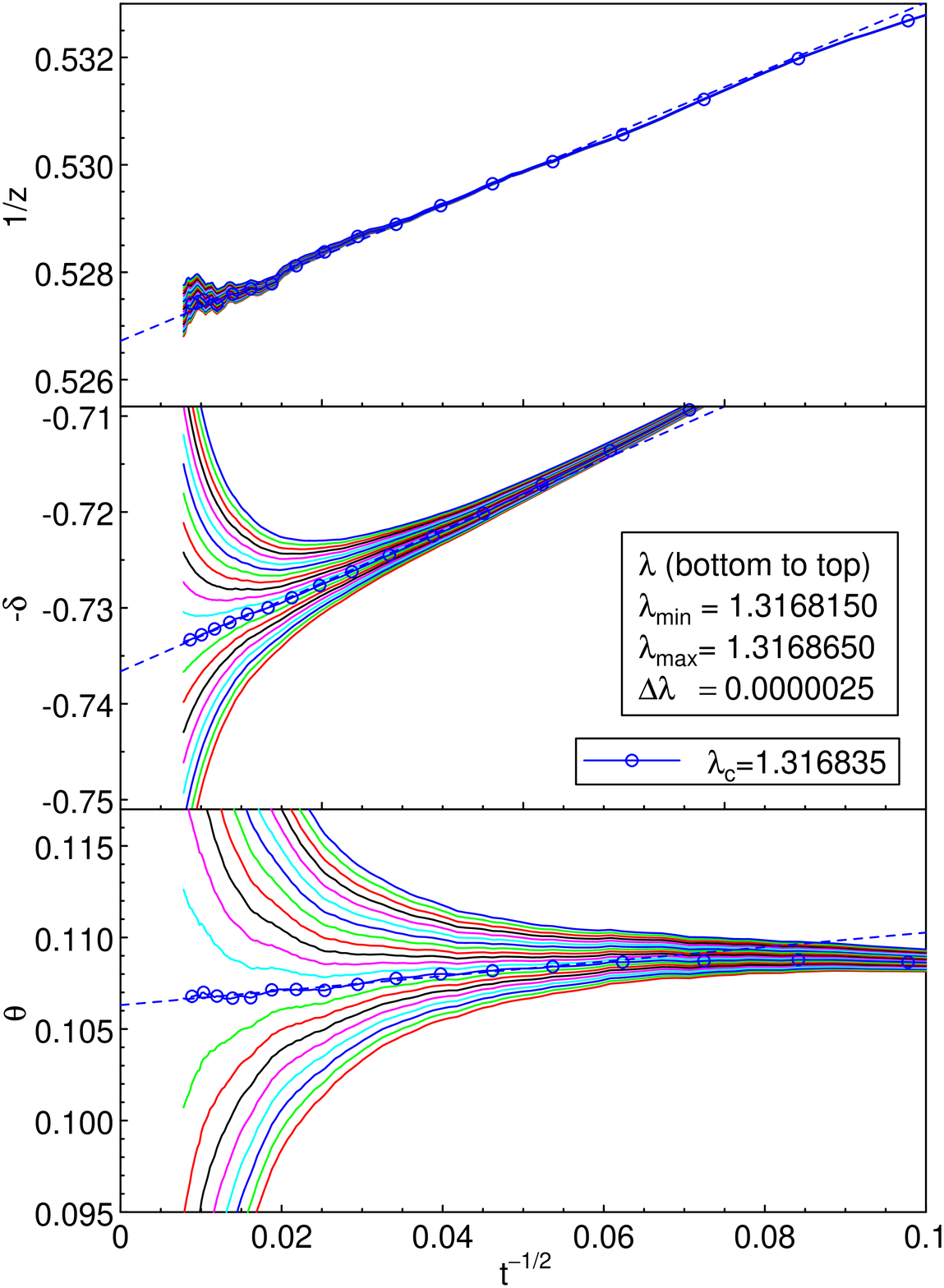}}
\caption{(Color online) Effective critical exponents $1/z= (d\ln R) / (d\ln t)$, $\delta = (d\ln P_s) / (d\ln t)$ and
      $\Theta = (d\ln N_s) / (d\ln t)$ vs. $t^{-1/2}$. The critical curves are marked by dots,  and the
      dashed lines are linear extrapolations to $t=\infty$. }
\label{fig:cl_exponents}
\end{figure}

To determine the exponents $z$, $\delta$, and $\Theta$, we apply the same type of analysis
to the logarithmic derivatives of $R$, $P_s$, and $N_s$ with respect to time. The corresponding graphs are shown
in Fig.\ \ref{fig:cl_exponents}.
Extrapolation to $t=\infty$ yields the dynamical exponent $1/z=0.5267(1)_{\rm sys}(1)_{\rm ran}$
as well as $\delta=0.7367(5)_{\rm sys}(1)_{\rm ran}$ and $\Theta=0.1062(2)_{\rm sys}(2)_{\rm ran}$.
These values fulfill hyperscaling because $\Theta + 2\delta -3/z =-0.0005(22)$, in excellent agreement
with the exact result of zero.

Finally, we measure the exponent combination $1/(\nu_\perp z)$ by analyzing the time dependencies of
$({\partial \ln P_s})/({\partial \lambda})$ and $({\partial \ln N_s})/({\partial \lambda})$ according to
(\ref{eq:dlnrho}). Extrapolating the effective exponent to $t=\infty$ as above yields
$1/(\nu_\perp z)=0.9040(5)_{\rm sys}(5)_{\rm ran}$. The correlation length and order parameter
exponents can be calculated by combining this value with our results for $z$ and $\beta/\nu_\perp$
yielding $\nu_\perp = 0.5826(9)$ and $\beta=0.815(2)$.

In Table \ref{tab:clean_exponents}, we compare our estimates for the critical exponents with
earlier results.
\begin{table}
\renewcommand*{\arraystretch}{1.2}
\begin{tabular}{cllllll}
\hline\hline
Value              & This work    & Ref.\ \cite{SanderOliveiraFerreira09} & Ref.\ \cite{Dickman99} & Ref.\ \cite{SabagOliveira02} & Ref.\ \cite{Jensen92}\\
\hline
$\lambda_c^0$        & 1.316835(1)  & (*)                                   & 1.31686(1)             & 1.31683(2)                   & 1.3168(1) \\
\hline
$\beta/\nu_\perp$  & 1.3991(4)    & 1.395(4)                              & {\it 1.394(1)}         &                              & {\it 1.392(5)}\\
$\nu_\perp$        & 0.5826(9)    &                                       & {\it 0.580(3)}         &                              & {\it 0.584(6)}\\
$\beta$            & 0.815(2)~~   &                                       & {\it 0.808(5)}         & 0.78(1)                      & 0.813(11) \\
$\delta$           & 0.7367(6)    &                                       & 0.7263(11)             &                              & 0.732(4) \\
$\Theta$           & 0.1062(4)    &                                       & 0.110(1)               &                              & 0.114(4) \\
$z$                & 1.8986(8)    & 1.916(5)                              & {\it 1.919(4)}         &                              & {\it 1.901(5)}\\
\hline
$2/z$              & 1.0534(4)    &                                       & 1.042(2)               &                              & 1.052(3) \\
$\nu_\perp z$      & 1.106(2)~~   &                                       & 1.114(4)               &                              & 1.11(1)  \\
$\Theta z$         & 0.2016(6)    & 0.216(3)                              &                        &                              & \\
$D_f$              & 1.6009(4)    &                                       &                        & 1.56(3) \\
\hline\hline
\end{tabular}
\caption{Critical infection rate and critical exponents of the clean three-dimensional
contact process. The upright numbers are directly given in the respective papers, the italic ones were calculated
using scaling relations.
The fractal dimension $D_f=3-\beta/\nu_\perp$. (*) The authors of Ref.\ \cite{SanderOliveiraFerreira09} used the value
of $\lambda_c$ found in Ref.\ \cite{Dickman99} as an input. }
\label{tab:clean_exponents}
\end{table}
The present estimates have significantly higher precision than the values in the literature. They are roughly compatible with Jensen's
values \cite{Jensen92} within their given errors (for $\Theta$, the difference is about twice the given error, though). However,
they are clearly not compatible with the values given in Refs.\ \cite{Dickman99} and \cite{SanderOliveiraFerreira09}
(for $\delta$, the difference is about ten times the given error, and for $z$ it is about five times the given error).
We believe, this discrepancy can be traced back to the location of the critical point. According to our data,
the infection rate $\lambda=1.31686(1)$, identified as critical in Ref.\  \cite{Dickman99} and also employed in
\cite{SanderOliveiraFerreira09}, is on the active side of the transition (it differs from our estimate by about
three times the given error). As the survival probability decays more slowly in the active phase than at criticality,
this may be responsible for the low $\delta$-value and, via the hyperscaling relation, for
the high $z$-value reported in Ref.\ \cite{Dickman99}.

\subsection{Contact process on a diluted lattice}
\label{subsec:MC_diluted}

The remainder of Sec.\ \ref{sec:MC} focuses on the contact process on a diluted
lattice. We tried to use the same reweighting technique as in the clean case
to save computer time. However, these attempts were not successful. 
The reweighing method of  Ref.\ \cite{Dickman99}
considers a set of simulation runs (particular realizations of the
Markov process) at some infection rate $\lambda$ and reweighs their statistical 
probabilities according to a slightly different $\lambda'$.
This only works as long as the two infection rates are sufficiently close such that
their sets of possible runs overlap significantly. This overlap decreases with increasing
simulation time. In the presence of disorder, particularly long simulation times are required 
because the critical dynamics is logarithmically slow. 
Thus, reweighting is restricted to very narrow $\lambda$-intervals
(too narrow compared to the range of infection rates we needed to explore to
determine the critical point).
All results were thus obtained in the
conventional manner by performing a separate run for each $\lambda$-value.

\begin{figure}
\centerline{\includegraphics[width=8.5cm]{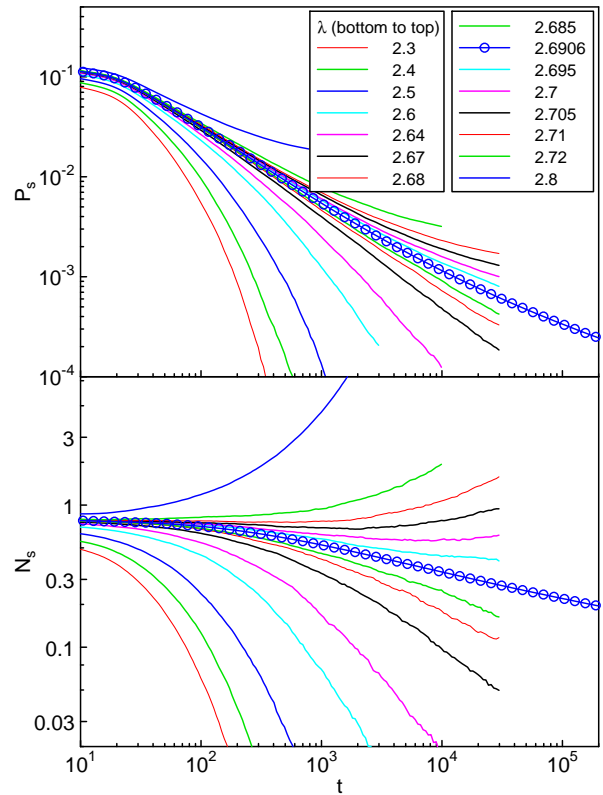}}
\caption{(Color online) Survival probability $P_s$ and number of active sites $N_s$ vs.
         time $t$ for impurity concentration $p=0.5$ and several infection rates $\lambda$.
         The critical curve at $\lambda_c=2.6906$ is marked by dots. }
\label{fig:p05_overview}
\end{figure}

Figure \ref{fig:p05_overview} gives an overview over spreading simulations (starting from
a single active seed site) for a vacancy concentration $p=0.5$.
The data represent averages over at least 5000 disorder configurations, with 128 trials
starting from random seed sites for each configuration. A system size of $500^3$ sites ensured
that the active cluster stayed smaller than the sample for the entire simulation run.
The figure shows that the dynamics in the vicinity of the phase transition is very slow. In particular,
the time-dependence of the survival probability appears to be slower than a power law, in agreement
with the activated scaling scenario of Sec.\ \ref{subsec:activated}. Moreover, the data show indications of Griffiths
singularities, i.e., nonuniversal power-law behavior somewhat below the critical infection rate.
We also note that the number of sites in the active cluster $N_s$ decreases with time at the transition,
in contrast to the clean case and to the diluted case in two dimensions \cite{VojtaFarquharMast09}.
This implies a negative exponent $\bar\Theta$.

To find the precise location of the critical point within the activated scaling scenario,
one might be tempted to search for power-law relations between $\ln t$ and observables such
$P_s$ and $N_s$ (either by plotting $\ln P_s$ and $\ln N_s$ vs. $\ln\ln t$ or by analyzing
the corresponding effective exponents). However, this method is highly unreliable as
the unknown microscopic time scale $t_0$ in (\ref{eq:Ps_t_activated}) and (\ref{eq:Ns_t_activated})
provides a correction to scaling via $\ln(t/t_0) = \ln t -\ln t_0$.
This strongly influences the results because the simulations cover only a moderate range in $\ln t$.
(In the two-dimensional simulations, Ref.\ \cite{VojtaFarquharMast09}, it was found that neglecting $t_0$
could change the apparent value of $\bar\delta$ from its correct value of 1.9 to 3.)

To  circumvent this problem, we follow the method devised in Ref.\ \cite{VojtaFarquharMast09}.
It is based on the observation that $t_0$ has the same value in the scaling forms of
all quantities because it is related to the energy scale $\Omega_0$ of the underlying SDRG. Consequently,
if one analyzes the relation between $N_s$ and $P_s$ or other such combinations of observables,
the critical point corresponds to power-law behavior (independent of the value of $t_0$)
as long as all other corrections to scaling are small.

\begin{figure}[tb]
\centerline{\includegraphics[width=8.5cm]{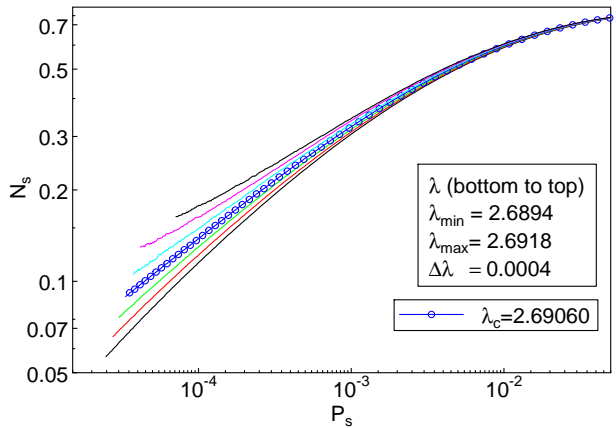}}
\caption{(Color online) $N_s$ versus $P_s$ for vacancy concentration
         $p=0.5$ and several infection rates
         $\lambda$ close to the critical point.}
\label{fig:p05_ns_ps}
\end{figure}

We performed long spreading runs with a maximum time of $5\times 10^7$, system size
$500^3$ and vacancy concentration $p=0.5$ for several infection rates $\lambda$
close to the phase transition. The resulting
plot of $N_s$ vs $P_s$ is shown in Figure \ref{fig:p05_ns_ps}.
The data are averages over $10^5$ to $10^6$ disorder configurations with 1000 trials starting
from random seed sites for each configuration. The figure shows that the relation between
$N_s$ and $P_s$ indeed approaches a power law in the long-time (small $P_s$) limit.
The figure also indicates that the crossover to the asymptotic behavior is very slow.
The asymptotic power law is only reached when $P_s$ falls well below $10^{-3}$ which
corresponds to times larger than $10^4$, implying that long simulations are required
to determine the critical behavior. Moreover, the mean-square radius of the active cluster
at the crossover time is approximately 25, implying a total diameter of about 100.
This means that simulations of systems with less than $100^3$ sites will never reach
the asymptotic critical behavior.

\subsection{Critical exponents}
\label{subsec:MC_critical}

\begin{figure}
\centerline{\includegraphics[width=8.5cm]{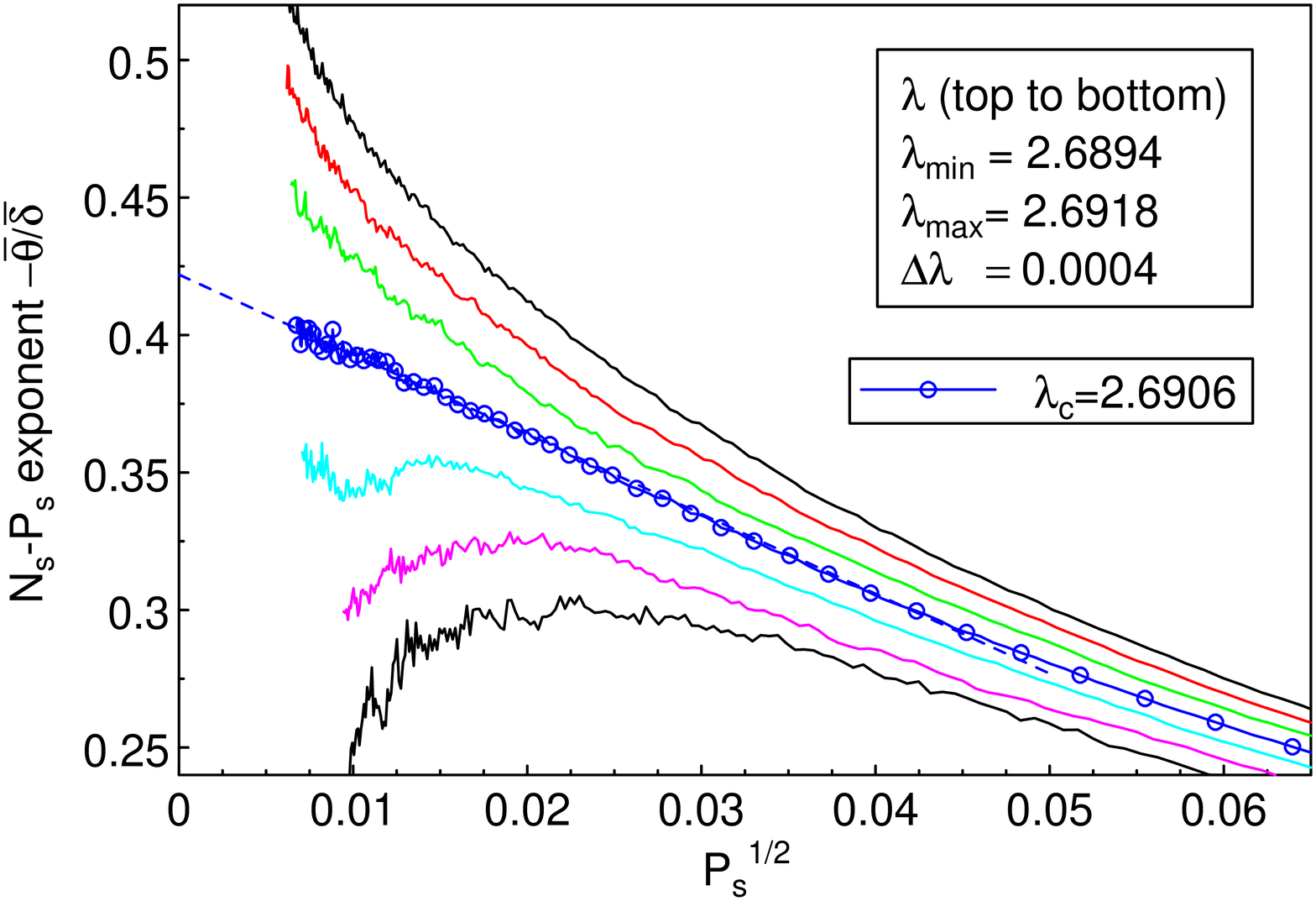}}
\caption{(Color online) Effective critical exponent $-\bar\Theta/\bar\delta = (d\ln N_s) / (d\ln P_s)$ vs.
         $P_s^{1/2}$ calculated from the data in Fig.\ \ref{fig:p05_ns_ps}.
         The critical curve is marked by dots, and the dashed line is a linear
         extrapolation to $P_s=0$. }
\label{fig:p05_nsps_exponent}
\end{figure}

To find the critical infection rate $\lambda_c$ and to measure the finite-size scaling exponent $\beta/\nu_\perp$
(the scale dimension of the order parameter), we define the effective (running) exponent
$(d\ln N_s) /(d\ln P_s) = -\bar\Theta/\bar\delta$.
It is related to the finite-size scaling exponent via $\bar\Theta/\bar\delta = 3\nu_\perp/\beta - 2$
[see eqs.\ (\ref{eq:Ps_t_activated}) and (\ref{eq:Ns_t_activated})].
To avoid the uncertainties stemming from the unknown microscopic time scale $t_0$, we extrapolate
this effective exponent to $P_s=0$ rather than $t=\infty$.
Figure \ref{fig:p05_nsps_exponent} shows the effective
exponent as a function of $P_s^{\bar y}$ with $\bar y=1/2$.
(The value 1/2 was again chosen empirically to permit an approximately linear extrapolation
of the data with $P_s \lesssim10^{-3}$. Because of the slow crossover to the
asymptotic regime, the value of $\bar y$ is much more uncertain than that of $y$ in the clean case,
see Fig.\ \ref{fig:cl_nsps_exponent}.)

From Fig.\ \ref{fig:p05_nsps_exponent}, we conclude that $\lambda_c=2.6906(3)$ for a vacancy concentration of $p=0.5$.
Extrapolating the effective exponent to $P_s=0$ yields $\bar\Theta/\bar\delta=-0.42(3)$
where the error estimate largely stems from the uncertainty in $\lambda_c$ (and the related uncertainty in
$\bar y$.) The statistical error is much smaller. The resulting finite-size scaling exponent is $\beta/\nu_\perp = 1.90(4)$.
An estimate for this exponent can also be obtained from analyzing the dependence of $N_s$ on
$R$ in a similar fashion. The data show additional curvature (corrections to scaling), thus giving
the less precise value $\beta/\nu_\perp = 1.85(15)$.

\begin{figure}
\centerline{\includegraphics[width=8.5cm]{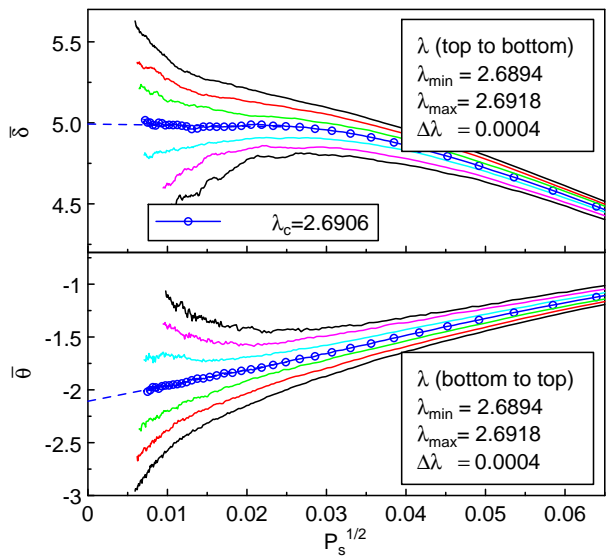}}
\caption{(Color online) Effective critical exponents $\bar\delta = -(d\ln P_s) / (d\ln \ln(t/t_0))$ and
          $\bar\Theta = (d\ln N_s) / (d\ln \ln(t/t_0))$ vs.
         $P_s^{1/2}$ for $p=0.5$ and $t_0=1$.
         The critical curves are marked by dots, and the dashed lines are linear
         extrapolations to $P_s=0$. }
\label{fig:p05_exponents}
\end{figure}

We now apply the same type of analysis to the logarithmic derivatives of $P_s$, $N_s$, and $R$
with respect to $\ln(t/t_0)$ to determine the values of the exponents $\bar\delta$, $\bar\Theta$, and $\psi$.
This requires a value for the microscopic time scale
$t_0$. Since an incorrect $t_0$ would produce additional corrections to scaling,
we estimated its value by minimizing the time dependence ($P_s$ dependence) of the effective
exponents $\bar\delta$ and $\bar\Theta$. This yields $t_0 \approx 1.0(4)$.
Figure \ref{fig:p05_exponents} shows the resulting effective exponents $\bar\delta$ and
$\bar\Theta$ as a function of $P_s^{1/2}$ for vacancy concentration $p=0.5$.
Extrapolating the data at the critical infection rate $\lambda_c=2.6906$ to $P_s=0$ (i.e., $t=\infty$)
gives the values $\bar\delta =5.0(2)$ and $\bar\Theta=-2.1(2)$. Again, the error estimate is dominated
by the uncertainty in $\lambda_c$ (and the resulting uncertainties in $\bar y$ and $t_0$).
The tunneling exponent $\psi$ can be determined by combining the value of $\beta/\nu_\perp$ with either
$\bar\delta$ or $\bar\Theta$. We find $\psi = 0.38(3)$. Alternatively, $\psi$ can be obtained from
the dependence of $R$ on $\ln(t/t_0)$. As these data show additional corrections to scaling, the
extrapolation to $t=\infty$ is difficult and leads to the less precise estimate $\psi=0.41(5)$.

To find the critical exponents $\nu_\perp$ and $\beta$, we now study the dependence of
$({\partial \ln P_s})/({\partial \lambda})$ and $({\partial \ln N_s})/({\partial \lambda})$
on $\ln(t/t_0)$ according to eq.\ (\ref{eq:dlnrho_activated}). This yields the exponent
combination $1/\nu_\perp \psi = 2.7(3)$. Combining this with the value for $\psi$, we obtain
$\nu_\perp=1.0(2)$. This analysis is hampered by the rather large uncertainties in $\psi$ and
$t_0$. A better estimate can be obtained by considering the dependence of
$({\partial \ln P_s})/({\partial \lambda})$ and $({\partial \ln N_s})/({\partial \lambda})$
on $P_s$ which takes the form
\begin{equation}
\frac{\partial \ln P_s}{\partial \lambda} \sim \frac{\partial \ln N_s}{\partial \lambda}
\sim P_s^{-1/\beta}~.
\end{equation}
Extrapolating the effective exponents to $P_s=0$ as before, we obtain the values
$1/\beta$ =0.53(2) and 0.55(3) from the $P_s$ and $N_s$ data, respectively. Our final estimate
of the order parameter exponent is thus $\beta=1.87(7)$. Combined with the finite-size scaling
exponent, this yields $\nu_\perp=0.98(6)$.

\begin{figure}
\centerline{\includegraphics[width=8.5cm]{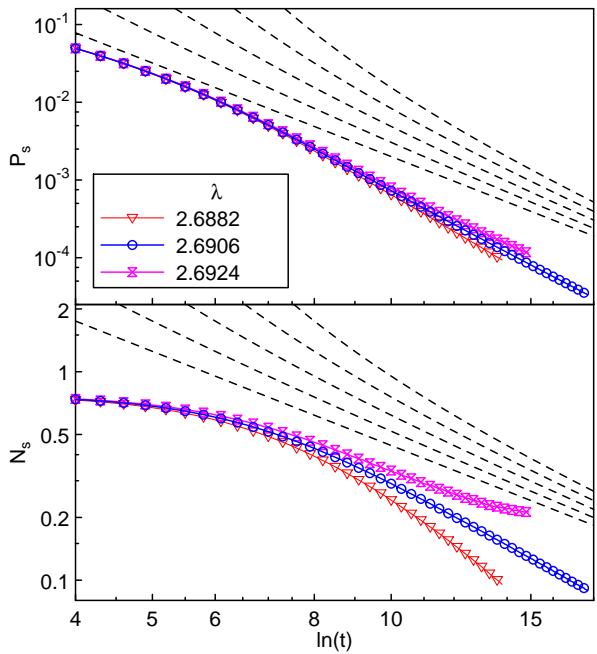}}
\caption{(Color online) Time dependence of $P_s$ and $N_s$ for impurity concentration $p=0.5$
and several $\lambda$ close to the transition compared to the predictions of the numerical SDRG of Ref.\
\cite{KovacsIgloi11}. The dashed lines represent the functions $P_s \sim \ln(t/t_0)^{-4}$ and $N_s \sim \ln(t/t_0)^{-1.5}$
with $\ln(t_0)$ = 0, 1, 2, 3, 4 (bottom to top) and arbitrary prefactor.}
\label{fig:p05_comparison}
\end{figure}

In Table \ref{tab:dirty_exponents}, we compare our estimates for the critical exponents with
results of a numerical SDRG calculation \cite{KovacsIgloi11} of the random transverse-field Ising model
with up to $128^3$ sites. This model is expected to be in the same universality class as the
disordered contact process.
\begin{table}
\renewcommand*{\arraystretch}{1.2}
\begin{tabular*}{7cm}{@{\extracolsep{\fill}}cll}
\hline\hline
Value              & This work    & Ref.\ \cite{KovacsIgloi11} \\
\hline
$\beta/\nu_\perp$  & 1.90(4)    & 1.84(2) \\
$\nu_\perp$        & 0.98(6)    & 0.99(2)       \\
$\beta$            & 1.87(7)    & {\it 1.82(4)}          \\
$D_f$              & 1.10(4)    & {\it 1.16(2)}          \\
\hline
$\psi$             & 0.38(3)    &   0.46(2)  \\
$\nu_\perp \psi$   & 0.37(4)    &  {\it 0.45(3)}         \\
$\bar\delta$       & 5.0(2)     &   {\it 4.0(2)}          \\
$\bar\Theta$       & -2.1(2)    &   {\it -1.5(1)}        \\
\hline\hline
\end{tabular*}
\caption{Critical exponents of the disordered three-dimensional contact process compared to results of
the SDRG calculation \cite{KovacsIgloi11}.
The upright numbers are directly given in Ref.\ \cite{KovacsIgloi11}, the italic ones were calculated
using scaling relations.
The fractal dimension $D_f=3-\beta/\nu_\perp$.}
\label{tab:dirty_exponents}
\end{table}
All static exponents (above the dividing line in the table) agree within their error bars
(though just barely in the case of $\beta/\nu_\perp$). In contrast, the tunneling exponent
$\psi$ and the other exponents characterizing the time dependencies (below the dividing
line) do not agree. This suggests that the uncertainties in determining the microscopic time scale
$t_0$ and, correspondingly, the microscopic energy scale $\Omega_0$ of the SDRG calculation
may be responsible for the disagreement because $t_0$ and $\Omega_0$ do not influence the static exponents.
We note, however, that our raw data do not seem to be compatible with the values for $\bar\delta$ and $\bar\Theta$
calculated from the results of Ref.\ \cite{KovacsIgloi11} even if we allow $t_0$ to vary
(unless one assumes a crossover from our $\bar\delta = 5$ to $\bar\delta=4$ and from
$\bar\Theta=2$ to $\bar\Theta=1.5$ at times $t \gtrsim 10^8$
beyond the range of our simulations).
This can be seen in Fig.\ \ref{fig:p05_comparison} where we compare our data to
the functions $P_s \sim \ln(t/t_0)^{-4}$ and $N_s \sim \ln(t/t_0)^{-1.5}$
with $\ln(t_0)$ = 0, 1, 2, 3, and 4.
We will return to this question in the concluding section.

\subsection{Universality of the critical behavior}
\label{subsec:MC_universality}

So far, all results on the disordered contact process were for a vacancy concentration $p=0.5$.
We now address the question of whether or not the critical behavior is universal, i.e., independent
of the disorder strength. The SDRG underlying the infinite-randomness scenario becomes exact
only for infinitely strong disorder (infinitely broad disorder distributions). Therefore, it cannot decide
the fate of a weakly disordered system. However, Janssen's perturbative renormalization group
\cite{Janssen97}, which is controlled for weak disorder, shows runaway flow towards large disorder
strength. Furthermore, Hoyos \cite{Hoyos08} showed that within an improved SDRG scheme, the disorder
always increases under renormalization, even if it is weak initially. These arguments support
a universal scenario in which the critical behavior is independent of the disorder strength.

\begin{figure}
\centerline{\includegraphics[width=8.5cm]{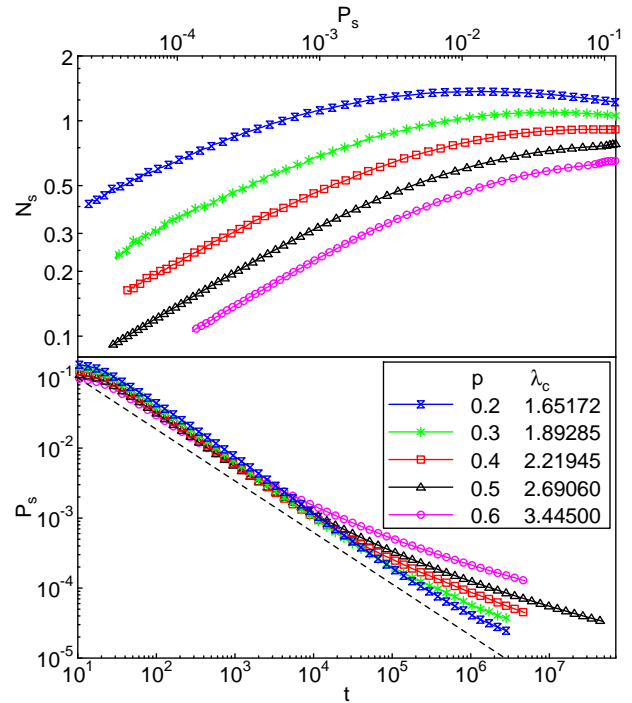}}
\caption{(Color online) Upper panel: $N_s$ vs. $P_s$ at criticality for several
    vacancy concentrations $p$.
    Lower panel: $P_s$ vs. $t$ at criticality, demonstrating the crossover from
    the clean to the dirty critical behavior. The dashed line represents a power law with
    the clean critical exponent $\delta=0.7366$ and arbitrary prefactor. }
\label{fig:universality}
\end{figure}

To study the question of universality numerically, we performed simulations for vacancy concentrations
$p=0.2$, 0.3, 0.4, and 0.6 in addition to the value 0.5. Repeating the complete analysis as discussed
in the previous subsections for all values of $p$ would have been prohibitively expensive in terms of
computer time. We therefore focused on finding the finite-size scaling exponent from  $N_s$ vs. $P_s$
plots analogously to Fig.\ \ref{fig:p05_ns_ps}, using somewhat shorter runs. The maximum time was
at least $3 \times 10^6$ for all vacancy concentrations, and the data are averages over
at least $10^7$ trials using systems of $500^3$ or $720^3$ sites. The resulting critical curves
are presented in the upper panel of Fig.\ \ref{fig:universality}.
In the low-$P_s$ (long-time) limit, all curves appear to be parallel, implying that
$\bar\Theta/\bar\delta$ and with it the finite-size scaling exponent $\beta/\nu_\perp$
takes the same value for all vacancy concentrations. The figure also suggests that
the weak-disorder curves (in particular $p=0.2$) have not fully crossed over to the asymptotic
critical behavior. This is confirmed in the lower panel of Fig.\ \ref{fig:universality}
which presents a log-log plot of $P_s$ vs. $t$ at criticality. While the stronger-disorder
curves ($p=0.5$, 0.6) start to deviate from the clean critical power law at $t \approx
10^3$ to $10^4$ (in agreement with the estimate discussed at the end of Sec.\
\ref{subsec:MC_diluted}), the $p=0.2$ curve deviates appreciably only after $t\approx 10^6$.
(Note that these long crossover times also imply huge system sizes to reach the asymptotic regime.
For $p=0.2$, the mean-square radius of the active cloud at the crossover time of $10^6$ is about
200.)

Our simulations thus show no indications of nonuniversal, continuously varying critical
exponents. However, we cannot rigorously exclude that the exponents change for very weak disorder,
because the extremely large crossover times between the clean and the dirty critical behavior
prevent us from reaching the asymptotic regime in these cases.

\subsection{Griffiths region}
\label{subsec:MC_griffiths}

\begin{figure}
\centerline{\includegraphics[width=8.5cm]{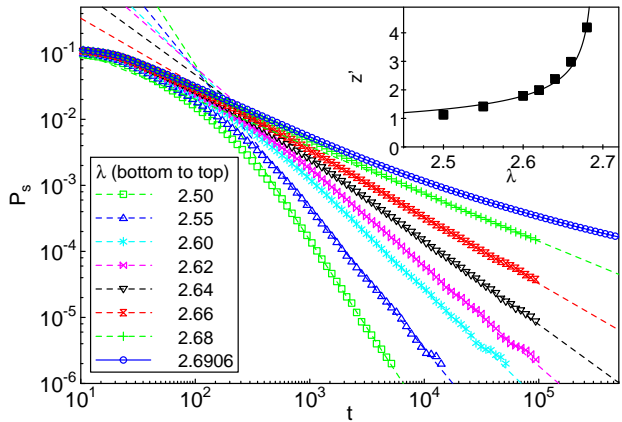}}
\caption{(Color online) $P_s$ vs. $t$ for several infection rates $\lambda$ in the Griffiths region
$\lambda_c^0<\lambda<\lambda_c$ (vacancy concentration $p=0.5$).
The dashed lines are power-law fits of the long-time behavior to
 (\ref{eq:griffithspower}). Inset: Resulting Griffiths dynamical exponent $z'$ as a function of
 $\lambda$. The solid line is a fit to (\ref{eq:z_prime}).}
\label{fig:p05_griffiths}
\end{figure}

In order to investigate the Griffiths region $\lambda_c^0<\lambda<\lambda_c$, we have also performed
detailed simulations for infection rates $\lambda$ below but close to the critical rate $\lambda_c$.
Figure \ref{fig:p05_griffiths}
presents the resulting survival probability $P_s$ as a function of time $t$ for vacancy concentration
$p=0.5$.
The data are averages over at least 10000 disorder configurations with 1000 trials
per configuration. The system size is $300^3$ sites. For all infection rates shown,
the long-time decay of the survival probability obeys (over several orders of magnitude in $P_s$ and/or $t$)
the nonuniversal power law predicted by the rare region arguments of Sec.\ \ref{subsec:griffiths}.

The Griffiths dynamical exponent $z'$ can be found by fitting the long-time decay to
(\ref{eq:griffithspower}). The resulting values, shown in the inset of Fig.\ \ref{fig:p05_griffiths}, demonstrate
that $z'$ diverges as $\lambda$ approaches the critical value $\lambda_c=2.6906$.
Fitting $z'$ to the expected power law (\ref{eq:z_prime}) gives
a value for the combination $\nu_\perp \psi$. The fit is not of particularly high quality,
but the resulting value, $\nu_\perp \psi=0.42(6)$, is in reasonable agreement with the value determined at
criticality.

\subsection{Contact process at the lattice percolation threshold}
\label{subsec:MC_percolation}

This subsection is devoted to the nonequilibrium phase
transition of the diluted contact process across the lattice percolation threshold $p_c$.
In the phase diagram shown in Fig.\ \ref{fig:pd}, this transition is marked
by the vertical line at $p_c$ between $\lambda_c^{-1}=0$ and the multicritical point.

The contact process on a diluted lattice close to the percolation threshold
can be understood by combining classical percolation theory with
the properties of the supercritical contact process on finite-size clusters
\cite{VojtaLee06,LeeVojta09}.
Although its behavior follows the activated scaling scenario described in
subsection \ref{subsec:activated}, the critical exponents of the percolation transition
differ from those of the generic transition discussed in the preceding sections.
Interestingly, they are completely determined by the values of the classical
lattice percolation exponents $\beta_c$ and $\nu_c$ which are known
numerically with high accuracy \cite{StaufferAharony_book91}.

According to Refs.\  \cite{VojtaLee06,LeeVojta09}, the order parameter exponent $\beta$ and
the correlation length exponent $\nu_\perp$ of the nonequilibrium phase transition are identical to the
corresponding lattice exponents, $\beta=\beta_c=0.417$ and $\nu_\perp=\nu_c=0.875$.
The tunneling exponent $\psi$ is given by the fractal dimension $D_c$ of the critical
lattice percolation cluster, $\psi=D_c=3-\beta_c/\nu_c=2.523$. As a result,
the critical exponent $\bar\delta$ takes a very small value,  $\bar\delta =\beta/(\nu_\perp\psi)=0.188$.

\begin{figure}
\centerline{\includegraphics[width=8.5cm]{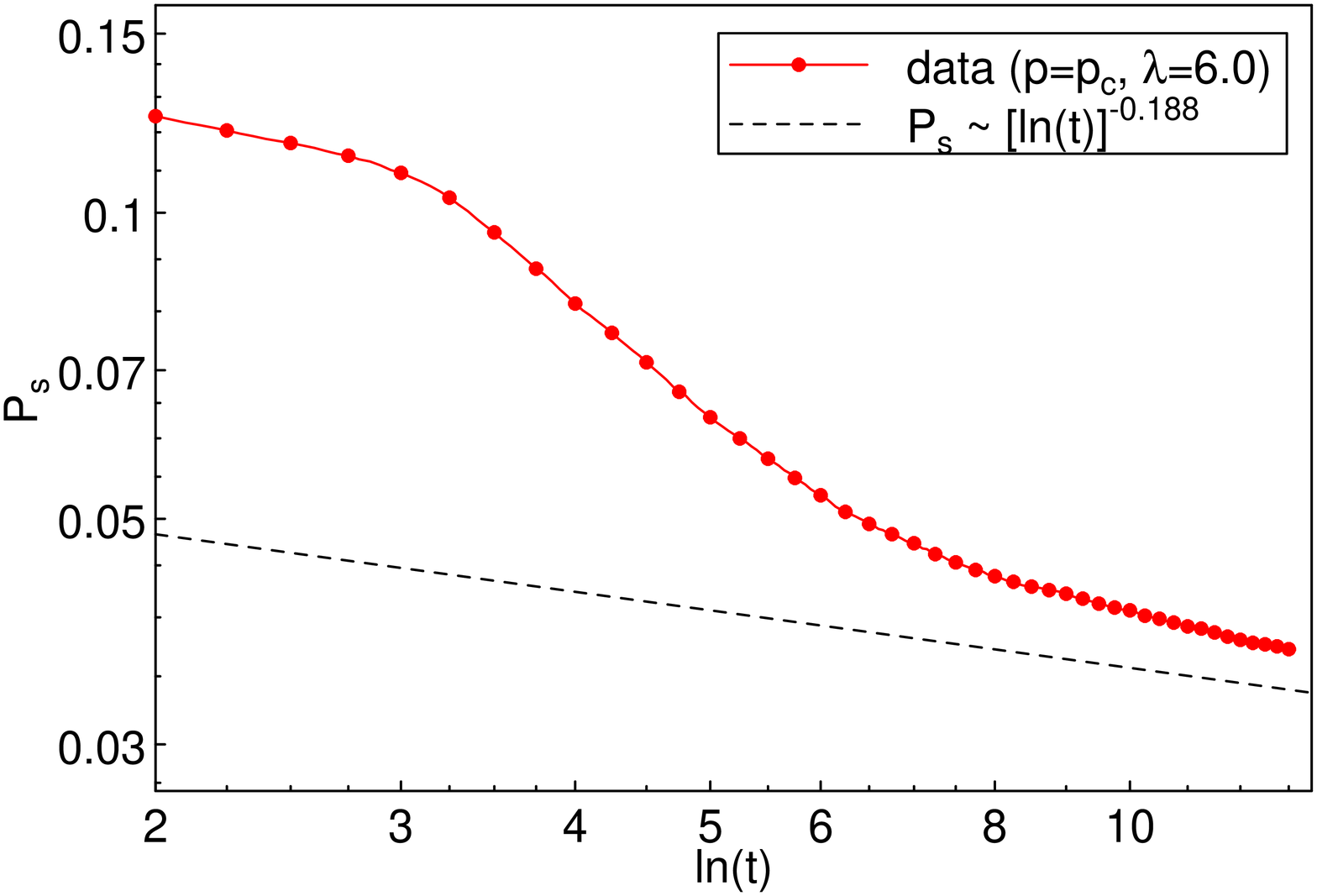}}
\caption{(Color online) $P_s$ vs $\ln(t)$ for $p=p_c=0.6883920$ and $\lambda=6.0$.
The dashed line represents the expected logarithmic long-time decay,
         $P_s \sim [\ln(t)]^{-\bar\delta}$ with $\bar\delta=0.188$ and arbitrary prefactor.  }
\label{fig:percolation}
\end{figure}

We performed spreading simulation runs (starting from a single active site)
at $p_c=0.6883920$ \cite{LorenzZiff98} and $\lambda=6.0$ to a  maximum time
of $5\times 10^5$. Because of the small value of $\bar\delta$, these simulations
are particularly time consuming. The resulting survival probability $P_s$ (averaged over 230
disorder configurations with 200 trials per configuration) is presented in Fig.\
\ref{fig:percolation}.
The data are in agreement with the qualitative predictions of Refs.\ \cite{VojtaLee06,LeeVojta09}:
a rapid initial decay towards a quasi-stationary state,
followed by a slow logarithmic time dependence due to the successive dying out of
the contact process on larger and larger lattice percolation clusters. The figure shows that
the long-time behavior of our data is compatible with the predicted exponent value
$\bar \delta =0.188$. However, we found it impossible to calculate a
precise value of $\bar \delta$ directly from the simulation data because
the long-time decay is extremely slow.

\section{Summary and conclusions}
\label{sec:conclusions}

To summarize, we performed large-scale Monte Carlo simulations of the contact process on
site-diluted cubic lattices. We determined the infection rate--dilution phase diagram.
It features two different nonequilibrium phase transitions, (i) the generic transition
that occurs for dilutions below the percolation threshold of the lattice and
is driven by the dynamic fluctuations of the contact process, and (ii) the transition
across the percolation threshold which is driven by the lattice geometry.

Our simulation results show that the generic transition is controlled by an infinite-randomness
critical point for all dilutions investigated. It gives rise to ultraslow activated
(exponential) dynamical scaling instead of the power-law dynamical scaling at conventional
critical points. The corresponding logarithmic time dependencies of various observables
at criticality required long simulation times and thus a huge numerical effort
(in total about 100\,000 CPU days on the Pegasus cluster at Missouri S\&T).
We determined the complete critical behavior of the generic transition
and found it to be universal, i.e., independent of the disorder strength (dilution).

The critical exponents
are listed in Table \ref{tab:dirty_exponents}, together with results of a numerical
SDRG calculation \cite{KovacsIgloi11} of the three-dimensional random
transverse-field Ising model which is predicted to be in the same universality class.
We were able to calculate reasonably accurate estimates for the static critical exponents
including the finite-size scaling exponent $\beta/\nu_\perp$, the order-parameter
exponent $\beta$ and the spatial correlation length exponent $\nu_\perp$.
The correlation length exponent satisfies the inequality $d\nu_\perp > 2$,
as is expected in a disordered system \cite{CCFS86}.
Our values for the static exponents agree with the
corresponding numerical SDRG results of Ref.\ \cite{KovacsIgloi11} within the given errors.
In contrast, our result for the tunneling exponent $\psi$  as well as $\bar\delta$ and $\bar\Theta$
do not agree with the values quoted in Ref.\ \cite{KovacsIgloi11}. Estimates
of $\psi$, $\bar\delta$ and $\bar\Theta$ depend sensitively on the microscopic time scale $t_0$
or, correspondingly, on the microscopic energy scale $\Omega_0$ in the numerical SDRG calculation,
while the static exponents are independent of it. This suggests that
uncertainties in the value of $t_0$ or $\Omega_0$ may be responsible
for the disagreement. However, even if we allow the value of $t_0$ to vary, our $P_s$ and
$N_s$ data do not seem to be compatible with the values of $\bar\delta$ and $\bar\Theta$
predicted by Ref.\ \cite{KovacsIgloi11}.

The differences between our results and the numerical SDRG calculation could either imply a real
difference in universality class, or they could mean that one or both sets of results represent
effective rather than true asymptotic exponents.
A full resolution of this question will likely require much more extensive simulations together
with a careful analysis of finite-size effects and, in particular, of the microscopic
time/energy scale in the infinite-randomness scenario.

In addition to the generic transition, we briefly studied the nonequilibrium transition across
the lattice percolation threshold. The Monte Carlo results support the predictions of the theory
developed in Refs.\ \cite{VojtaLee06,LeeVojta09}: The dynamical critical behavior is of activated
type with critical exponents that are combinations of the classical lattice percolation
exponents. Our simulations also allowed us to find with reasonable accuracy the location  of the
multicritical point separating the generic transition from the percolation transition (see Fig.\ \ref{fig:pd}).
However, as the dynamics is expected to be even slower than that of the generic transition,
finding the true multicritical behavior appears to be beyond our current computational
resources.

We also obtained high-precision estimates for the critical behavior of the
three-dimensional DP universality class by performing simulations of the \emph{clean}
contact process in three dimensions. We employed the reweighting technique proposed by
Dickman \cite{Dickman99} to save computer time. By using large lattices of up to $999^3$ sites
and long times of up to $5 \times 10^4$, we were able to compute the critical exponents
with unprecedented accuracy (see Table \ref{tab:clean_exponents}). As the dynamics of the clean contact process is much faster
than that of the disordered contact process, this part of the work took only a small fraction
 (about 5000 CPU days) of the overall computer time.

Let us now put our results into broader perspective. Our results for the critical behavior
of the disordered contact process in three dimensions are in agreement with
a general classification \cite{VojtaSchmalian05,Vojta06}
of phase transitions in quenched disordered systems
according to the effective dimensionality $d_{\rm eff}$ of the defects and the
lower critical dimension $d_c^-$ of the problem.
(A) If $d_{\rm eff} < d_c^-$, the critical point is of conventional power-law type and
accompanied by exponentially weak Griffiths singularities.
In class (B), which contains systems with $d_{\rm eff}=d_c^-$,
the critical behavior is controlled by an infinite-randomness
fixed point with activated scaling, accompanied by strong power-law Griffiths singularities.
(C) For  $d_{\rm eff}>d_c^-$, the rare regions can undergo the phase
transition independently from the bulk system. This leads to a destruction of the sharp
phase transition by smearing \cite{Vojta03a}.
For the contact process with vacancies (point defects), $d_{\rm eff}=d_c^-=0$
leading to class B. In contrast, the contact process with
\emph{extended} (line or plane) defects belongs to class C \cite{Vojta04,DickisonVojta05}.

We conclude by noting that the exotic critical behavior of the disordered contact process
(in one, two, and three dimensions) may be responsible for the striking absence of
directed percolation scaling in at least some of the experiments \cite{Hinrichsen00b}.
In view of the increased experimental activities in the area of absorbing-state transitions
\cite{TKCS07,CCGP08,FFGP11,OkumaTsugawaMotohashi11}, we hope that our theoretical results
will help guiding the data analysis in further experiments.
However, it must be pointed out that the extremely slow dynamics and narrow critical region
will prove to be a challenge for the verification of the activated scaling scenario not
just in simulations but also in experiments.
Finally, we emphasize that our results are of importance beyond absorbing state transitions.
The strong-disorder renormalization group predicts our transition to belong to a broad
universality class that also includes, e.g., the three-dimensional random transverse-field Ising
model \cite{Fisher92,Fisher95,IgloiMonthus05}. Consequently, the critical behavior found here
should be valid for other systems in this universality class as well.

\section*{Acknowledgements}

This work has been supported in part by the NSF under grants no. DMR-0906566 and
DMR-1205803.

\bibliographystyle{apsrev4-1}
\bibliography{../00Bibtex/rareregions}
\end{document}